# Car following model simulating traffic breakdown and concave growth pattern of oscillations in traffic flow

Zuojun Wang[1], Junfang Tian[1*], Rui Jiang[2*], Xiaopeng Li[3], Shou Feng Ma[1]

[1]Institute of Systems Engineering, College of Management and Economics, Tianjin University, No. 92 Weijin Road, Nankai District, Tianjin 300072, China

[2]MOE Key Laboratory for Urban Transportation Complex Systems Theory and Technology, Beijing Jiaotong University, Beijing 100044, China

[3]Department of Civil and Environmental Engineering, University of South Florida, FL, USA 33620

Traffic breakdown, as one of the most puzzling traffic flow phenomena, is characterized by sharply decreasing speed, abruptly increasing density and in particular suddenly plummeting capacity. In order to clarify its root mechanisms and model its observed properties, this paper proposes a car-following model based on the following assumptions: (i) There exists a preferred time-varied and speed-dependent space gap that cars hope to maintain; (ii) there exists a region $R$ restricted by two critical space gaps and two critical speeds in the car following region on the speed-space gap diagram, in which cars' movements are determined by the weighted mean of the space-gap-determined acceleration and the speed-difference-determined acceleration; and (iii) out of region $R$, cars either accelerate to the free flow speed or decelerate to keep safety. Simulation results show that this model is able to simultaneously reproduce traffic breakdown and the transition from the synchronized traffic flow to wide moving jams. To our knowledge, this is the first car-following model that is able to fully depict traffic breakdown, spontaneous formation of jams, and the concave growth of the oscillations.

# 1. Introduction

Traffic breakdown from free flow to congested flow and spontaneous formation of jams are two of the most complex highway traffic flow phenomena that have long puzzled traffic researchers. Traffic breakdown, whose mechanisms and quantitative properties are arguably the oldest and most central question of traffic flow theory (Schönhof and Dirk Helbing, 2009), are characterized by a sharp decrease in speed, an abrupt increase in density, and in particular a plummeting drop in capacity (Agyemang-Duah and Hall, 1991; Cassidy and Bertini, 1999). The identification of traffic breakdown can be traced back to the finding of the reverse-lambda shaped structure of the flow-density data (Edie and Foote, 1958; Edie, 1961). The tip of the "lambda" corresponds to high-flow states of free traffic, which are not stable and may rapidly drop into congested flow marked by the other leg of the "lambda" and thus cause traffic breakdown.

---

[*]Corresponding author.
Email: jftian@tju.edu.cn (JFT), jiangrui@bjtu.edu.cn (RJ).

Persaud and Hurdle (1988) reported that traffic breakdown can be induced by an upstream bottleneck. Koshi (1986) and Mahnke et al. (2005) found that the occurrence of traffic flow breakdown at freeway bottlenecks is strongly related to platoons in the bottleneck area. Elefteriadou et al. (1995) discovered that traffic breakdown has the probabilistic nature, i.e. it may occur or not at the flows that are lower than the maximum observed flow. Lorenz et al. (2000) and Okamura et al. (2000) found that breakdown probability increases with increasing flow rate. Brilon et al. (2005) conducted a series studies on German freeways traffic flow data, and proposed an empirical approach to analyze the probability of breakdown based on univariate Weibull distribution with respect to flow. In order to prevent traffic breakdown, various control strategies such as variable speed control and ramp metering (Chow et al., 2009) have been proposed and utilized in real traffic.

Spontaneous formation of jams, which is also known as "phantom traffic jams", refers to formation of jams with no obvious reason such as an accident or a bottleneck. The phenomenon was firstly reported by Treiterer and Myers (1974), in which vehicle trajectories were obtained by arterial photography. The traffic experiment on a circular road (Sugiyama et al., 2008) has further demonstrated that jams form spontaneously when the traffic density is large enough.

To simulate traffic breakdown from free flow to congested flow and spontaneous formation of jams, many traffic flow models have been proposed. In traditional traffic flow models, the mechanisms of the two phenomena are the same. In these models, it is assumed that there exists a unique relationship between speed and density (or spacing) in the steady state. The steady traffic flow is unstable or metastable in certain density range. Small disturbances are able to grow in unstable or metastable traffic, which leads to traffic breakdown and spontaneous formation of jams (see Treiber and Kesting, 2013; Laval and Leclercq, 2010; Chen et al. 2014).

However, these traditionally models have been questioned in the last two decades. In particular, based on the empirical data (Kerner and Rehborn, 1996a, 1996b, 1997; Kerner, 1998, 2004, 2009), Kerner claimed that the congested flow can be further classified into synchronized flow and wide moving jam. Thus, there are three phases in traffic flow: Free flow (F), Synchronized flow (S) and Jam (J). Accordingly there are three kinds of phase transitions in traffic flow: transitions from F to S (F→S), S to J (S→J), and F to J (F→J). The transition F→J seldom happens unless the formation of SF is strongly hindered due to non-homogeneity, in particular at a traffic split on a highway (Kerner, 2000). The traffic breakdown from free flow to congested flow thus corresponds to F→S, and spontaneous formation of jams corresponds to S→J. Kerner explained the occurrence of F→S as a competition between speed adaptation and over-acceleration and S→J as a pinch effect.

The debate about Kerner's three-phase traffic theory and the traditional traffic theory is mainly due to a lack of high-fidelity traffic data. Recently, we have organized experimental study of car following behavior on an open road section (Jiang et al., 2014, 2015). The location and speed of each individual car have been recorded by high precision GPS devices. Thus, high-fidelity traffic data concerning the complete evolution of the disturbances have been obtained.

Our experiment found that (i) the spacing between a leading car and a following car can change significantly even though the speeds of the two cars are essentially constant and the speed difference is very small; (ii) platoon length might be significantly different even if the average speed of the platoon is essentially the same; (iii) the standard deviation of oscillations grows in a concave way along the platoon. These findings run against the basic assumption of traditional linear car-following models, in which the standard deviation of oscillations initially grows in a convex way. Later, the concave growth of traffic oscillations are demonstrated by the empirical trajectory data (Tian, et al., 2016).

Our experiment findings demonstrate that the traffic states span a 2D region in the speed-spacing (or density) plane. We have proposed two possible mechanisms to produce this feature. (i) In a certain range of spacing, drivers are not so sensitive to the changes in spacing when the speed differences between cars are small. Only when the spacing is large (small) enough, will they accelerate (decelerate) to decrease (increase) the spacing. (ii) At a given speed, drivers do not have a fixed preferred spacing.

We have proposed two car-following models based on the two mechanisms, respectively. The proposed models are able to reproduce the spontaneously jam formation and the concave growth of oscillations. Nevertheless, both models fail to describe the traffic breakdown from free flow to synchronized traffic flow (i.e., F→S transition). Motivated by this fact, this paper proposed a new car-following model, which is able to simultaneously reproduce the F→S transition and the S→J transition. To our knowledge, this is the first car-following model that is able to fully depict traffic breakdown, spontaneous formation of jams, and the concave growth of oscillations.

The remainder of the paper is organized as follows. Section 2 presents the model. Section 3 shows that the model can reproduce the concave growth of oscillations consistent with field observations. Section 4 investigate the properties of the model with simulation results of fundamental diagrams and spatiotemporal diagrams in the standard test scenario of circular road and discusses the contribution of space gap related acceleration and speed difference related acceleration in the model. In section 5, simulations on an open road with different bottlenecks are conducted. Finally section 6 concludes the paper.

## 2. Car-following model

In order to construct a car following model that is able to reproduce traffic breakdown and the other associated traffic flow phenomena raised in the previous section, the following assumptions are proposed.

**Assumption 1:** There exists a preferred space gap denoted by desired space gap $d_{n,\text{de}}$, that car $n$ hopes to maintain. We allow this gap to be time-variant and dependent on vehicle speed.

This assumption is to describe the experimental observation in Jiang et al (2015) that the spacing between two consecutive cars can change significantly even though the speeds of the two cars are almost identical and approximately remain constant.

**Assumption 2**: There exists a region $R$ restricted by two critical space gaps ($d_{\text{sa}}<d_n<d_{\text{fr}}$) and two critical speeds ($v_c<v_n<v_{\max}$) in the car following region, where $d_{\text{sa}}$ is the safe space gap, $d_{\text{fr}}$ is the freely moving space gap, and $v_c$ is the lower threshold, $v_{\max}$ denotes the maximum speed. See Figure 1 for illustration.

This assumption distinguishes the car following behavior in the high speed state ($v_c<v_n$) from that in the low speed state ($v_c<v_n$). This assumption aims to reproduce traffic breakdown and the associated synchronized traffic flow. Empirical observations indicate that traffic breakdown happens in the free flow speed and meanwhile triggers the emergence of the synchronized traffic flow that can steadily exist when the speed of the traffic flow is greater than some critical value. When the speed of traffic flow is lower than the critical value, wide moving jams will occur in the synchronized traffic flow.

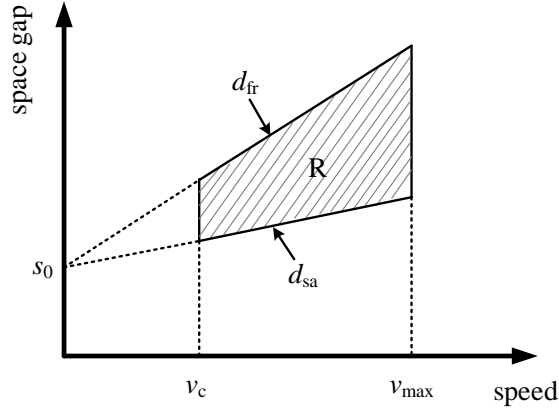

**Figure 1**. The space gap-speed plane where the dash area denotes the region R restricted in $d_{sa}<d_n<d_{fr}$ and $v_c<v_n<v_{max}$.

**Assumption 3**: In region $R$, the movement of car $n$ is determined by the weighted mean of the space-gap-determined acceleration and the speed-difference-determined acceleration. By adjusting their weights, we can explore the effects of space gap and speed difference on the traffic flow evolution.

This assumption is different from the traditional traffic flow theory that assumes the existence of the speed-space gap curve (or the fundamental diagram). It is also different from the three-phase traffic flow theory that presumes cars only react to the speed difference in a two-dimensional speed-space region. We believe this assumption is consistent with our daily experiences. The simulation results demonstrate that the space gap determined acceleration plays a decisive role in reproducing capacity drop.

**Assumption 4**: Out of region $R$ ($d_{sa}>d_n$ or $d_n<d_{fr}$ or $v_c>v_n$), the cars either accelerate to the free flow speed or decelerate to keep safety. Therefore, either free flow maintains or the phase transition from synchronized traffic flow to wide moving jams happens.

With the consideration of these assumptions, the car following model is formulated as follows. Consider a platoon of consecutive cars indexed by $n = 1,2,\cdots,N$ from upstream to downstream following each other on a single-lane highway. Without loss of generality, we investigate the car-following behavior of a generic car $n$ following its predecessor car $n+1$. The proposed car-following model assumes that there is a region $R$ in the space gap-speed plane. The driving behavior in this region is different from that out of the region, see Figure 1. This region is bounded by

$$d_{sa}<d_n<d_{fr} \text{ and } v_c<v_n<v_{max} \tag{1}$$

Here, $v_n$ is speed of car $n$, $d_n=x_{n+1}-x_n-L_{car}$ is space gap of car $n$, where $x_{n+1}$ and $x_n$ are respectively the location of the leading car and the following one, $L_{car}$ is the car's length.

$$d_{sa} = \max\left(v_n T_{sa} - \frac{v_n \Delta v_n}{2\sqrt{ab}},\ 0\right) + s_0 \tag{2}$$

$$d_{fr} = \max\left(v_n T_{fr} - \frac{v_n \Delta v_n}{2\sqrt{ab}},\ 0\right) + s_0 \tag{3}$$

here $s_0$ is bumper-to-bumper distance in jam. $T_{sa}$ and $T_{fr}$ respectively denote the safe driving time gap (lower threshold) and upper threshold of time gap; $b$ is assumed to decrease with speed,

$$b = b_{max} - (b_{max} - b_{min}) \cdot \frac{v_n}{v_{max}} \tag{4}$$

Here $b_{max}$ and $b_{min}$ are two deceleration parameters.

The acceleration of a car is determined as follows

$$a_n = \begin{cases} a_n^{IR}, & \text{in region } R, \\ a_n^{OR}, & \text{out of region } R. \end{cases} \tag{5}$$

*2.1 Acceleration in the region R*

The acceleration in the region $R$ is calculated by

$$a_n^{IR} = \begin{cases} a \cdot H(\lambda_1, \lambda_2), & \text{if } H(\lambda_1, \lambda_2) > 0, \\ b \cdot H(\lambda_1, \lambda_2), & \text{else} \end{cases} \tag{6}$$

with

$$H(\lambda_1, \lambda_2) = \alpha \cdot \lambda_1 + (1-\alpha) \cdot \lambda_2 \tag{7}$$

$$\lambda_1 = \begin{cases} -\dfrac{d_n - d_{n,de}}{d_{sa} - d_{n,de}}, & d_{sa} < d_n < d_{n,de} \\ \dfrac{d_n - d_{n,de}}{d_{fr} - d_{n,de}}, & d_{n,de} < d_n < d_{fr} \end{cases} \tag{8}$$

$$\lambda_2 = \begin{cases} 1, & \Delta v_n > \gamma v_n \\ \dfrac{\Delta v_n}{\gamma v_n}, & -\gamma v_n \leq \Delta v_n \leq \gamma v_n \\ -1, & \Delta v_n < -\gamma v_n \end{cases} \tag{9}$$

$$d_{n,de} = \max\left(v_n T_{n,de} - \frac{v_n \Delta v_n}{2\sqrt{ab}},\ 0\right) + s_0 \tag{10}$$

$$T_{n,de}(t + \Delta t) = \min\left(\max\left(T_{n,de}(t) + \xi,\ T_{sa}\right),\ T_{fr}\right) \tag{11}$$

Here $T_{n,de}$ is a preferred time gap that changes over time, $\xi$ is a uniformly distributed random number between $-\delta$ and $\delta$, where $\delta$ is a constant, $\gamma$ is a sensitivity parameter, $\Delta t$ is time step and is set to $0.1s$ in the model. $\lambda_1$ denotes space gap determined acceleration component, $\lambda_2$ denotes

speed difference determined acceleration component. Both components are in the range between
–1 and 1.

*2.2. Acceleration out of the region R*

The acceleration out of the region *R* is calculated by the Improved IDM model (Treiber and Kesting, 2013):

$$a_n^{\text{OR}} = \begin{cases} a\left(1-\left(\dfrac{v_n}{v_{\max}}\right)^4\right)\left(1-\left(\dfrac{d_{n,\text{de}}}{d_n}\right)^2\right), & \text{If } d_{n,\text{de}} \leq d_n, \\ a\left(1-\left(\dfrac{d_{n,\text{de}}}{d_n}\right)^2\right), & \text{Else} \end{cases} \quad (12)$$

## 3. Vehicle Platoon Calibration and Validation

We verify the concave growth of standard deviation of speed in the model with car-following experimental data collected by Jiang et al. (2014). In the experiments, the leading car of the platoon was asked to move with the speed $v_{\text{leading}}$ = 50, 40, 30, 15, and 7km/h, respectively, followed by another 24 cars. Within each $v_{\text{leading}}$, several runs were carried out, and stationary data are obtained to calculate the speed standard deviation. Then the results were averaged over the runs. Jiang's experimental results are illustrated with blue dotted line in Figure 1. One can see the concave growth of the speed standard deviation.

We applied the standard deviation of speed under $v_{\text{leading}}$ = 50, 30 and 7km/h to calibrate the model and data under $v_{\text{leading}}$ = 40 and 15km/h for validation. The Root Mean Square Percentage Error (*RMSPE*) is used as the fitness function.

$$RMSPE_\sigma = \sqrt{\dfrac{1}{24}\sum_{n=2}^{25}\left(\dfrac{\sigma_{v,n}^{\text{simu}} - \sigma_{v,n}^{\text{exp}}}{\sigma_{v,n}^{\text{exp}}}\right)^2} \quad (13)$$

where *n* denotes the car number, $\sigma_{v,n}^{\text{simu}}$ ($\sigma_{v,n}^{\text{exp}}$) represents the standard deviation of car *n* in the stationary state in the simulations (experiments).

The results of calibrated parameters are listed in Table 1. The calibration and validation results are shown in Table 2. All *RMSPE*s under different $v_{\text{leading}}$ are quite small, and the average RMSNE of calibration is 0.172 versus 0.167 resulting from validation.

We compare the simulation results with the experimental ones in Figure 2 and 3. From Figure 2, it can be seen that the simulated speed standard deviation also increases in a concave way in the platoon, which agrees with the experimental results quite well. Figure 3 shows that the stripe structures in the simulated spatiotemporal patterns also agree with the experimental ones.

**Table 1.** The calibrated parameter values

| Parameter | $a$ | $b_{min}$ | $b_{max}$ | $s_0$ | $v_{max}$ | $\delta$ | $\gamma$ | $v_c$ | $\alpha$ | $T_{sa}$ | $T_{fr}$ | $L_{car}$ |
|---|---|---|---|---|---|---|---|---|---|---|---|---|
| Value | 1.0 | 1.0 | 2.5 | 2 | 120 | 0.25 | 0.06 | 14.5 | 0.5 | 0.6 | 1.8 | 5 |
| Unit | m/s$^2$ | m/s$^2$ | m/s$^2$ | m | km/h | s | \ | m/s | \ | s | s | m |

**Table 2.** The calibration and validation with RMSPE of the new model

| | Calibration | | | Validation | |
|---|---|---|---|---|---|
| $v_{leading}$ (unit: km/h) | 7 | 30 | 50 | 15 | 40 |
| *RMSPE* | 0.200 | 0.152 | 0.164 | 0.260 | 0.074 |
| Average *RMSPE* | 0.172 | | | 0.167 | |

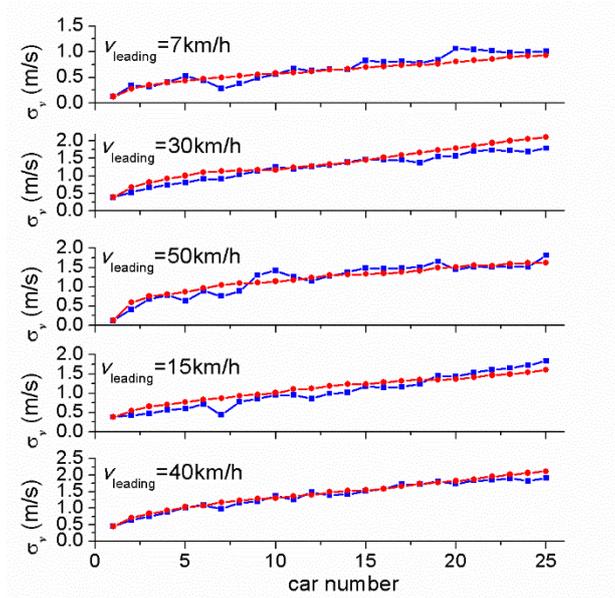

**Figure 2**. Simulated results (symbol solid red lines) of the car-following model versus experiment results (symbol solid blue lines) with the standard deviation of the speed for the cars. Car number 1 denotes the leading car.

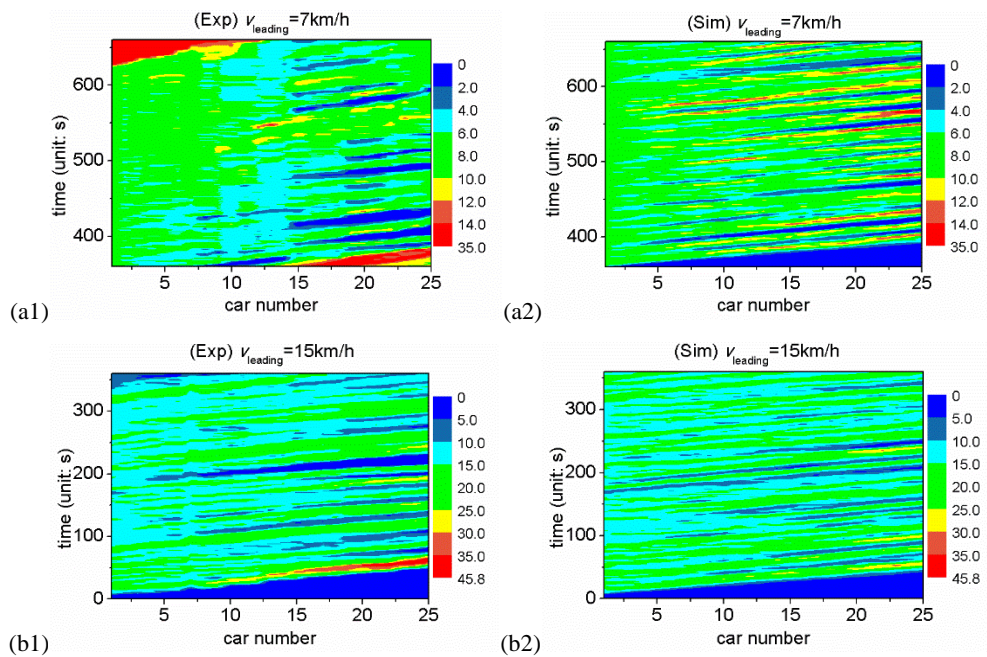

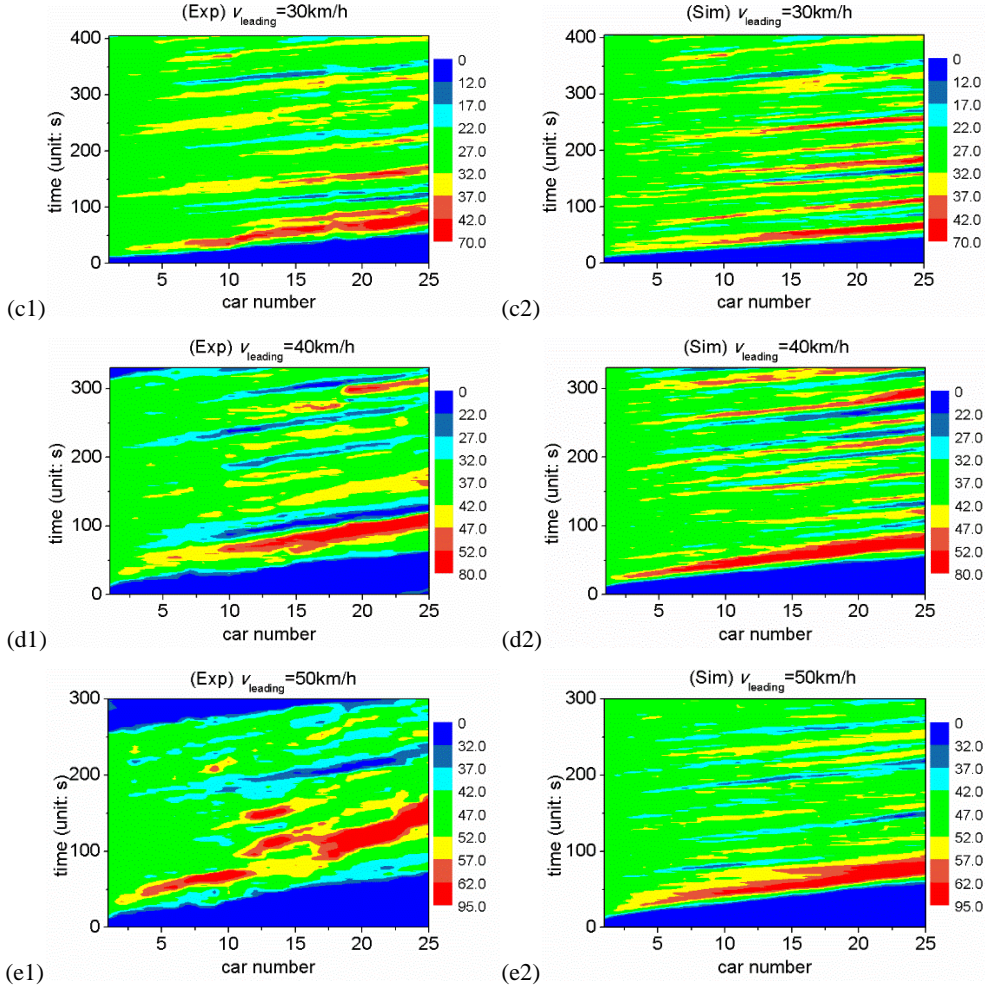

**Figure 3.** The spatiotemporal patterns of the platoon traffic. (a1-e1) are experimental results and (a2-e2) are simulation results. The color bar indicates speed (km/h).

## 4. General properties of the model

In this section, standard tests on a circular single lane road are conducted to investigate the properties of the model. As usual, two different initial car distributions are considered: 1) all cars are distributed homogeneously on the road; 2) all cars are distributed in a mega-jam. The parameters value used are shown in Table 1.

The flow-density and speed-density diagrams are shown in Figure 4. The empirically discovered reverse-lambda shaped structure of the flow-density diagram can be observed. The black dotted line represents the free flow state, the red dotted line represents the synchronized flow state, and the blue dotted line represents the wide moving jam state. The free flow states and the synchronized traffic flow states are obtained from the initial homogenous distribution. The wide moving jam states are obtained from the initial mega-jam distribution.

In the density range $k_1<k<k_2$, where $k_1$ and $k_2$ are critical density bound for traffic breakdown, traffic breakdown from free flow to synchronized flow can be observed, see Figure 5(a) for a typical example. One can see that the free flow is maintained for about 5 minutes. Then the traffic breakdown occurs and traffic flow evolves into the state that free flow and synchronized flow coexist. With the increase of density, the synchronized flow region expands and the free flow region shrinks, see Figure 6(a)[1]. When the density further increases, the synchronized flow becomes unstable and the transition to traffic jams is observed, see Figure 5(b). However, if the traffic starts from the initial mega-jam distribution, traffic flow will evolve into the coexist state of free flow and wide moving jams, see Figure 6(b).

Furthermore, we study the traffic breakdown probability from free flow to synchronized traffic flow, which is defined as follows. At each flow rate, we perform 200 runs of the simulations. The simulation time interval is set as 600s in each run. We denote the number of runs that traffic breakdown happens as $N_{br}$. Thus, traffic breakdown probability is calculated by $N_{br}/200 \times 100\%$. Figure 7 shows the result that the breakdown probability is also a monotonous increasing function of flow rate.

Consequently, the empirical characteristics of traffic breakdown, the spontaneous transitions from free flow to the synchronized traffic flow and from the synchronized traffic flow to wide moving jams, the concave growth pattern of oscillations, the metastable states, and the reverse-lambda shaped structure of the flux-density diagram have been successfully reproduced by the model.

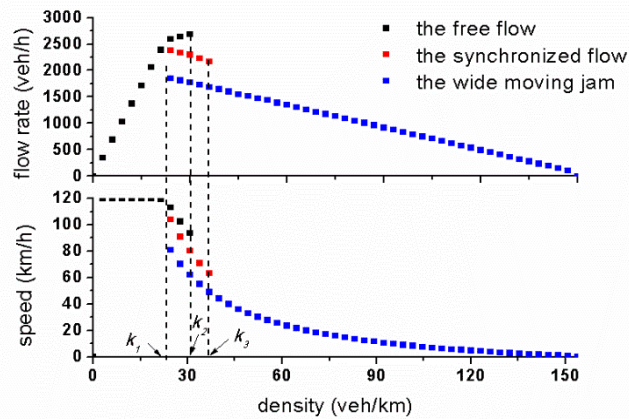

**Figure 4.** Flow-density and speed-density diagrams. The black dotted line represents the free flow, the red dotted line represents the synchronized flow, and the blue dotted line represents the wide moving jam. Fig. 4 describes the average flow of mixed-phase traffic and the average speed of all cars as the functions of the global density (number of cars divided by the circumference).

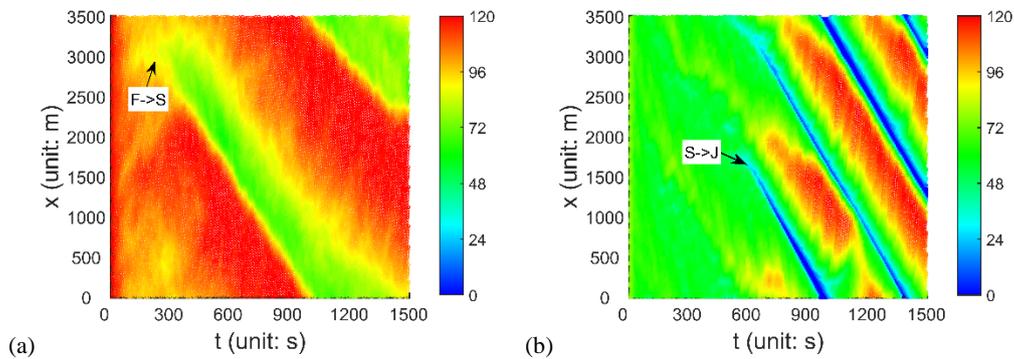

**Figure 5**. Simulation results of the traffic breakdown from the free flow to the synchronized flow (F→S) and the transition from the synchronized flow to the wide moving jam (S→J) on a circular road. The simulation starts from

---

[1] Note that at the density, when starting from a mega jam, the wide moving jam state will be observed, see Figure 6(b).

initial homogenous distribution with density (a): $k = 24$veh/km and (b): $k = 37$veh/km, respectively. The color bar indicates speed (unit: km/h).

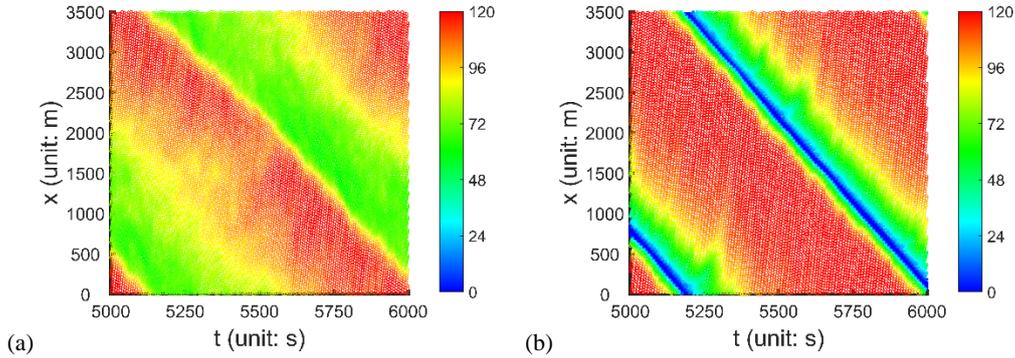

**Figure 6.** Simulation results of the spatiotemporal diagrams on a circular road with same density $k=26$veh/km under different initial distribution: (a) homogeneous; (b) mega-jam. The color bar indicates speed (km/h).

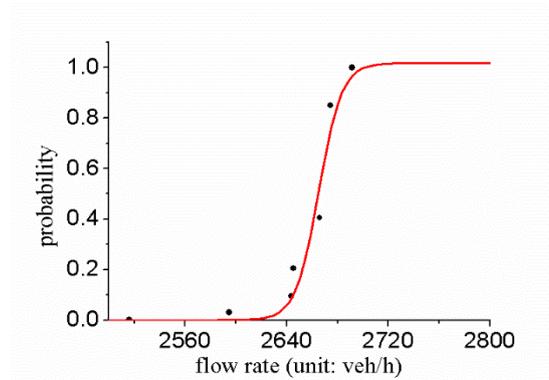

**Figure 7**. The plot of the breakdown probability versus the flow rate. The fitting logic function is
$$y = a/(1 + e^{-k(x-x_c)})$$ with $a=1.018$, $x_c=2665$, $k=0.111$.

Finally we discuss the contribution of space gap effect $\lambda_1$ and speed difference effect $\lambda_2$ to the phase transition behavior of traffic flow in the model. To this end, we change the value of weight factor $\alpha$ from 0 to 1. Figure 8 shows two typical results. When $\alpha$ decreases, the contribution of space gap effect $\lambda_1$ weakens. As a result, traffic breakdown phenomenon cannot be reproduced, see Figure 8(a) and (c). On the other hand, when $\alpha$ increases, the contribution of space gap effect $\lambda_1$ strengthens. Under the circumstance, the synchronized flow is not stable. Shortly after traffic breakdown from free flow to synchronized flow occurs, jam spontaneously emerges from the synchronized flow, see Figure 8(b) and (d). Therefore, the simulation demonstrates to depict the traffic flow realistically, both components of the space gap effect and the speed difference effect should be modeled properly.

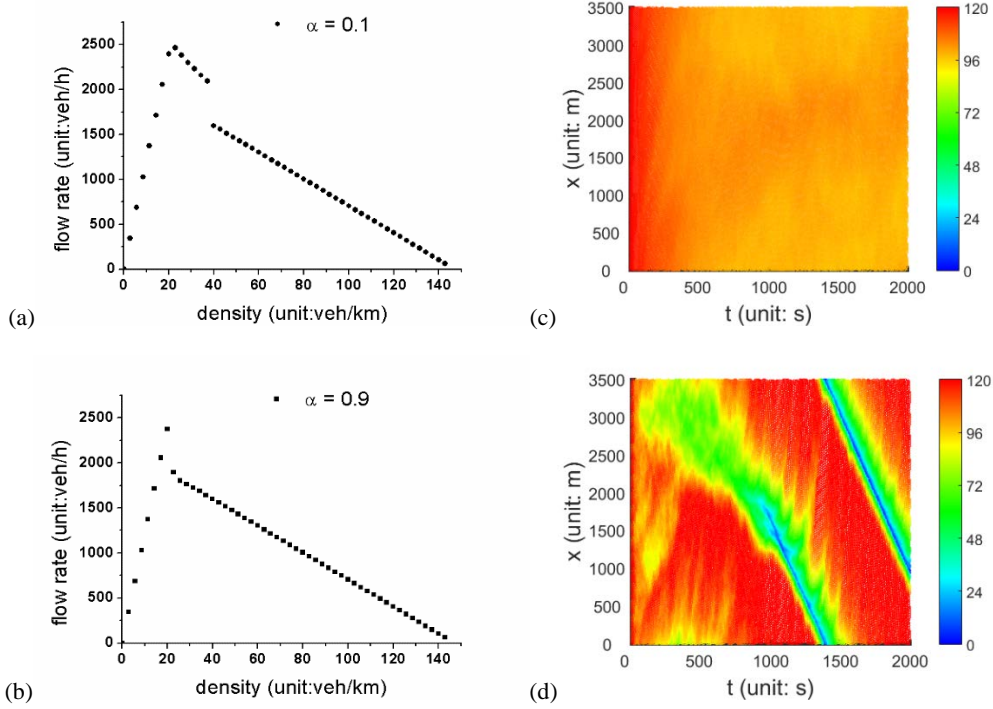

**Figure 8**. The flow-density diagrams and spatiotemporal diagrams with different values of α. (a, c) α=0.1, (b, d) α=0.9. In (c, d), the density k=24veh/km.

## 5 Traffic breakdown simulations

Since empirical findings show that traffic breakdown always happens in the bottleneck system, this section aims to demonstrate that it can be successfully reproduced by the car-following model. To show the generality of traffic breakdown, three kinds of bottlenecks are simulated respectively, i.e. the slope bottleneck, the speed limit bottleneck and the rubbernecking bottleneck.

Initially, the road with the length $L_{road}$ =112km is assumed to be filled with cars uniformly distributed with the density $k$ and speed $v_{max}$[2]. For the leading car, it will be removed when going beyond $L_{road}$. Its follower becomes the new leading one and moves freely. The bottleneck zone is located at [$0.8L_{road}$, $0.8L_{road}+L_{bottleneck}$]$m$ with $L_{bottleneck}$=300$m$. The rules of these bottlenecks are set as follows:

(i) The slope bottleneck. Under the circumstance, each car in the bottleneck zone would be affected by the gravity. Thus the acceleration is updated by

$$a_n(t) = a_n^M(t) - \kappa G \tag{14}$$

where $a_n^M(t)$ is the acceleration calculated by the model and $G = 9.81 m/s^2$ is the gravity acceleration. Here $\kappa$ denotes slope of the bottleneck zone and is set to 1.1% in the simulation.

---

[2] We would like to mention that when the density is below a threshold $k_1$, the traffic speed in the stationary state is very close to the maximum speed, see Figure 5.

(ii) The speed limit bottleneck. If the car in the bottleneck zone has a speed higher than speed limit $v_{lim}$, then the car decelerates with the deceleration $b_{lim}$ until the speed lowers down to $v_{lim}$. In the simulation, the parameters are set as $v_{lim} = 90km/h$ and $b_{lim}=2.5m/s^2$.

(iii) The rubbernecking bottleneck. When cars enter the rubbernecking zone, at each simulation time step, they have a probability $\gamma$ to rubberneck which will cause their speeds to decrease instantaneously by $\varphi\%$. Rubbernecking can occur at most once for each car in this zone. In the simulation, the parameters are set as $\gamma=0.1$, $\varphi=1.5$.

In the simulation, three virtual detectors are respectively placed at the location $x_{upstream1}=$ $(0.8L_{road}-100)m$, $x_{downstream2}= (0.8L_{road}+400)m$ and $x_{downstream3}= (0.8L_{road}+1200)m$ to collect the traffic data.

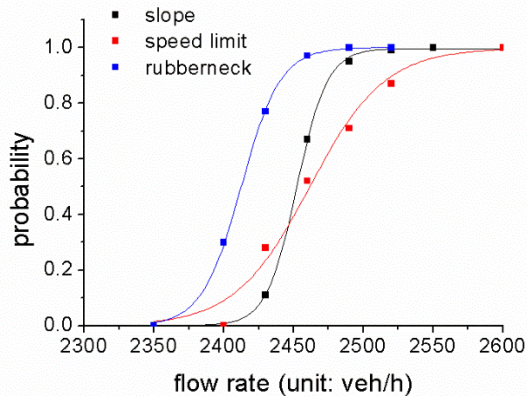

**Figure 9**. The plots of the traffic breakdown probability versus flow rate. The fitting logic functions $y = a/(1 + e^{-k(x-x_c)})$ with $a=0.997$, $x_c=2452$, $k=0.092$ for black curve, $a=0.998$, $x_c=2462$, $k=0.037$ for red curve, and $a=1.0$, $x_c=2413$, $k=0.071$ for blue curve.

Figure 9 shows that the breakdown probability is a monotonous increasing function of the flow rate, which is consistent with the empirical observation. Figure 10(a) shows a typical example of the traffic breakdown induced by the rubbernecking bottleneck. It can be seen that after this breakdown, the widening synchronized flow pattern (WSP) forms, which is also consistent with the empirical findings (Kerner, 2009). The measured time series of flow rate and speed from the three virtual detectors are presented in Figure 10(b). One can see that the free flow with flow rate $2400veh/h$ is maintained for about 15 mins before traffic breakdown occurs. When traffic breakdown happens, the free flow transits into the synchronized traffic flow, where the speed drops to about $75km/h$ and flow rate drops to about $2250veh/h$. At downstream detector 2, the recovering flow can be observed. At downstream detector 3, the traffic flow has already recovered to free flow.

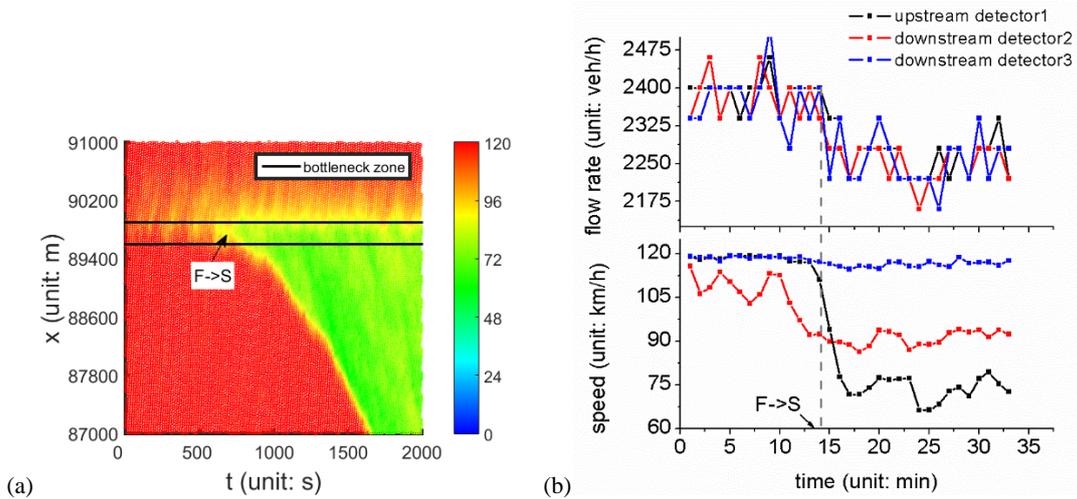

(a)  (b)

**Figure 10.** (a) The spatiotemporal diagram of traffic breakdown from free flow to congested flow on an open road with a rubberneck bottleneck. The color bar indicates speed (unit: km/h). (b)The time series of flow rate and speed measured by the three virtual detectors.The "upstream detector 1", "downstream detector 2", and "downstream detector 3" represent the detectors respectively at the upstream and downstream around the bottleneck region.

## 6. Conclusion

Traffic breakdown is characterized by sharply decreasing speed, abruptly increasing density and in particular suddenly plummeting capacity. It is arguably the oldest and most central question of traffic flow theory (Schönhof and Dirk Helbing, 2009). Many traffic flow models have been proposed to simulate the phenomenon. Kerner empirically found that traffic breakdown corresponds to a first order transition from the free flow to the synchronized flow, and spontaneous formation of jams corresponds to the transition from the synchronized flow to the wide moving jam phase (Kerner, 2009). These phenomena however cannot be reproduced by traditional car following models.

This paper proposed a car-following model based on the following assumptions: (i) There exists a preferred time-varied speed-dependent space gap that cars hope to maintain; (ii) there exists a region $R$ restricted by two critical space gaps and two critical speeds in the car following region, in which cars' movements are determined by the weighted mean of the space gap determined acceleration and the speed difference determined acceleration; (iii) out of region $R$, the cars either accelerate to the free flow speed or decelerate to keep safety. Through the calibration and validation by the experimental platoon data, it is shown that the concave growth of oscillations is well reproduced by the model. Simulation results show that empirical features of traffic breakdown, including the monotonous increasing function of traffic breakdown probability versus the flow rate, the emergence of the synchronized traffic flow, and the reverse-lambda shaped structure of the flux-density diagram, have been successfully reproduced. Furthermore, it is shown that to depict the traffic flow realistically, both components of the space gap effect and the speed difference effect should be modeled properly.

## Acknowledgements

JFT was supported by the National Natural Science Foundation of China (Grant No. 71431005, 71401120).                                                                 ☐ RJ was supported